\documentclass{elsart}

\usepackage{graphicx,epsfig,epsf,amssymb}
\usepackage{amssymb}

\newcommand{\dg}{\ensuremath{^\circ}}
\newcommand{\hess}{H.E.S.S.}
\newcommand{\RMS}{{\small RMS}}
\newcommand{\ADC}{{\small ADC}}
\newcommand{\PMT}{{\small PMT}}
\newcommand{\PMTs}{{\small PMT}s}
\newcommand{\NSB}{{\small NSB}}
\newcommand{\ARS}{{\small ARS}}
\newcommand{\ARSs}{{\small ARS}s}

\begin{document}

\begin{frontmatter}

\title{Calibration of cameras of the \hess\ detector}

\author[HD]{F. Aharonian}
\author[yerevan]{A.G. Akhperjanian} 
\author[durham]{K.-M. Aye}
\author[TO]{A.R. Bazer-Bachi}
\author[hamburg]{M. Beilicke}
\author[HD]{W. Benbow}
\author[HD]{D. Berge}
\author[PCC]{P. Berghaus}
\author[berlin,HD]{K. Bernl\"ohr}
\author[HD]{O. Bolz}
\author[meudon]{C. Boisson}
\author[berlin]{C. Borgmeier}
\author[berlin]{F. Breitling}
\author[durham]{A.M. Brown}
\author[durham]{P.M. Chadwick}
\author[LPNHE,LEA]{V.R. Chitnis\thanksref{india}}
\author[LLR]{L.-M. Chounet}
\author[hamburg]{R. Cornils}
\author[HD,LEA]{L. Costamante}
\author[LLR]{B. Degrange}
\author[potchefstroom]{O.C. de Jager}
\author[PCC]{A. Djannati-Ata\"{\i}}
\author[dublin]{L.O'C. Drury}
\author[berlin]{T. Ergin}
\author[PCC]{P. Espigat}
\author[gam]{F. Feinstein}
\author[LLR]{P. Fleury}
\author[LLR]{G. Fontaine}
\author[HD]{S. Funk}
\author[gam]{Y.A. Gallant}
\author[LLR]{B. Giebels}
\author[HD]{S. Gillessen}
\author[CEA]{P. Goret}
\author[LPNHE]{J. Guy}
\author[durham]{C. Hadjichristidis}
\author[HD2]{M. Hauser}
\author[hamburg]{G. Heinzelmann}
\author[grenoble]{G. Henri}
\author[HD]{G. Hermann}
\author[HD2,HD]{J. Hinton}
\author[HD]{W. Hofmann}
\author[potchefstroom]{M. Holleran}
\author[HD]{D. Horns}
\author[HD2,HD]{I. Jung}
\author[HD]{B. Kh\'elifi\corauthref{add1}}
\author[berlin]{Nu. Komin}
\author[berlin,HD]{A. Konopelko}
\author[durham]{I.J. Latham}
\author[durham]{R. Le Gallou}
\author[LLR]{M. Lemoine}
\author[PCC]{A. Lemi\`ere}
\author[LLR]{N. Leroy}
\author[berlin]{T. Lohse}
\author[TO]{A. Marcowith}
\author[HD,LEA]{C. Masterson}
\author[durham]{T.J.L. McComb}
\author[LPNHE]{M. de Naurois}
\author[durham]{S.J. Nolan}
\author[durham]{A. Noutsos}
\author[durham]{K.J. Orford}
\author[durham]{J.L. Osborne}
\author[LPNHE,LEA]{M. Ouchrif}
\author[HD]{M. Panter}
\author[grenoble]{G. Pelletier\thanksref{IUF}}
\author[PCC]{S. Pita}
\author[bochum]{M. Pohl\thanksref{iowa}}
\author[HD2,HD]{G. P\"uhlhofer}
\author[PCC]{M. Punch}
\author[potchefstroom]{B.C. Raubenheimer}
\author[hamburg]{M. Raue}
\author[LPNHE]{J. Raux}
\author[durham]{S.M. Rayner}
\author[LLR,LEA]{I. Redondo\thanksref{Sheffield}}
\author[bochum]{A. Reimer}
\author[bochum]{O. Reimer}
\author[hamburg]{J. Ripken}
\author[LPNHE]{M. Rivoal}
\author[charles]{L. Rob}
\author[LPNHE]{L. Rolland\corauthref{add2}}
\author[HD]{G. Rowell}
\author[yerevan]{V. Sahakian}
\author[grenoble]{L. Sauge}
\author[berlin]{S. Schlenker}
\author[bochum]{R. Schlickeiser}
\author[bochum]{C. Schuster}
\author[berlin]{U. Schwanke}
\author[bochum]{M. Siewert}
\author[meudon]{H. Sol}
\author[site]{R. Steenkamp}
\author[berlin]{C. Stegmann}
\author[LPNHE]{J.-P. Tavernet}
\author[PCC]{C.G. Th\'eoret}
\author[LLR,LEA]{M. Tluczykont}
\author[potchefstroom]{D.J. van der Walt}
\author[gam]{G. Vasileiadis}
\author[LPNHE]{P. Vincent}
\author[potchefstroom]{B. Visser}
\author[HD]{H.J. V\"olk}
\author[HD2]{S.J. Wagner}

\thanks[india]{now at Tata Institute of Fundamental Research, Homi Bhabha Road,
Mumbai~400~005, India}
\thanks[IUF]{Institut Universitaire de France}
\thanks[iowa]{now at Department of Physics and Astronomy,
Iowa State University, Ames, Iowa~50011-3160, USA}
\thanks[Sheffield]{now at Department of Physics and Astronomy, Univ. of Sheffield, The Hicks Building,
Hounsfiled Road, Sheffield~S3~7RH, U.K.}
 
\address[berlin]{
Institut f\"ur Physik, Humboldt Universit\"at zu Berlin, Newtonstr.~15,
    D-12489~Berlin, Germany}

\address[bochum]{
Institut f\"ur Theoretische Physik, Lehrstuhl~IV: Weltraum und Astrophysik,
    Ruhr-Universit\"at Bochum, D-44780~Bochum, Germany}
    
\address[dublin]{
Dublin Institute for Advanced Studies, 5~Merrion~Square, Dublin~2, Ireland}

\address[durham]{
University of Durham, Department of Physics, South Road, Durham DH1~3LE, U.K.}

\address[CEA]{
Service d'Astrophysique, DAPNIA/DSM/CEA, CE~Saclay, F-91191~Gif-sur-Yvette, France}

\address[grenoble]{
Laboratoire d'Astrophysique de Grenoble, Universit\'e Joseph Fourier, B.P.~53, F-38041~Grenoble~Cedex~9,
 France}

\address[hamburg]{
Universit\"at Hamburg, Inst. f\"ur Experimentalphysik, Luruper~Chaussee~149, D-22761~Hamburg, Germany}

\address[HD2]{
Landessternwarte, K\"onigstuhl, D-69117~Heidelberg, Germany}

\address[HD]{
Max-Planck-Institut f\"ur Kernphysik, P.O.~Box~103980, D-69029~Heidelberg, Germany}

\address[meudon]{
Observatoire de Paris, INSU/CNRS, Section de Meudon, Place~J.~Janssen,
F-92195~Meudon, France}

\address[gam]{
Groupe d'Astroparticules, IN2P3/CNRS, Universit\'e Montpellier~II, CC~85, Place Eug\`ene Bataillon,
 F-34095~Montpellier~Cedex~5, France}

\address[LLR]{
Laboratoire Leprince-Ringuet, IN2P3/CNRS,
Ecole~Polytechnique, F-91128~Palaiseau Cedex, France}

\address[LPNHE]{
Laboratoire de Physique Nucl\'eaire et de Hautes Energies, IN2P3/CNRS, Universit\'es Paris VI~\&~VII,
 4~place~Jussieu, F-75252~Paris~Cedex~05, France}

\address[PCC]{
Physique Corpusculaire et Cosmologie, IN2P3/CNRS, Coll{\`e}ge de France, 11~place~Marcelin~Berthelot,
 F-75231~Paris~Cedex~05, France}

\address[potchefstroom]{
Unit for Space Physics, North-West~University, Potchefstroom~2520,
    South~Africa} 

\address[charles]{
Institute of Particle and Nuclear Physics, Charles University, 
    V~Holesovickach~2, 180~00~Prague~8, Czech~Repulic}

\address[TO]{
Centre d'Etude Spatiale des Rayonnements CESR~-~CNRS/UPS, 9 Av. du Colonel Roche,
F-31029~Toulouse~Cedex, France}

\address[site]{
University of Namibia, Private~Bag~13301, Windhoek, Namibia}

\address[yerevan]{
 Yerevan Physics Institute, 2~Alikhanian Brothers St., 375036~Yerevan, Armenia}

\address[LEA]{
 European Associated Laboratory for gamma-ray astronomy, jointly 
supported by CNRS and MPG}

\corauth[add1]{Corresponding author, khelifi@mpi-hd.mpg.de}
\corauth[add2]{Corresponding author, rolland@lpnhep.in2p3.fr}

\begin{abstract}

\hess\ -- the High Energy Stereoscopic System-- is a new system of large atmospheric Cherenkov telescopes for
GeV/TeV astronomy. Each of the four telescopes of 107~$\mathrm{m^2}$ mirror area is equipped with a
960-pixel  photomulitiplier-tube camera. This paper describes the methods used to convert the photomultiplier
signals into the quantities needed for Cherenkov image analysis.  Two independent calibration techniques have
been applied in parallel to provide an estimation  of uncertainties. Results on the long-term stability of
the \hess\ cameras are also presented.

\end{abstract}

%

\end{frontmatter}

\section{Introduction}
\label{intro}

The Imaging Atmospheric Cherenkov Technique is the primary method for observations of  very high energy
($>$100~GeV) $\gamma$-rays. In this technique $\gamma$-rays are detected indirectly via the Cherenkov light
emitted by the charged particles of the induced air-shower.  Images of the shower in one or several
telescopes are analysed to provide background suppression and to reconstruct primary $\gamma$-ray parameters.

The \hess\ (High Energy Stereoscopic System) detector \cite{hofmann_tsukuba} consists of four telescopes
located in Namibia at 1800 meters altitude. Each telescope is composed of a  107~$\mathrm{m^2}$ mirror
\cite{optics1,optics2}  and a camera whose field of view is 5\dg\ in diameter. The camera focal plane is
covered by 960 photomultiplier tubes (\PMTs) with 0.16\dg\ angular extent. To decrease the camera read-out
window and the level of noise, very compact electronics located just behind the \PMTs\ is used (see \cite{pascal}
for a short description). The first \hess\ telescope has been operating since July 2002 and the system was
completed in December 2003.

It is essential for the extraction of parameters characterizing the air-shower images based on the raw  \PMT\ 
data to calibrate accurately the \PMTs\ and the electronic response. The calibration scheme  described below is
based in part on techniques developed for the CAT detector \cite{barrau},  the Durham~Mark~6 detector
\cite{mark6} and the HEGRA detector \cite{daum,gerd}. 

The parameters used in the  calibration of the \hess\ cameras and the methods used to derive them are
described below.  Results on the stability of photomultiplier gain and overall telescope efficiency are also
described.


\section{The \hess\ cameras}

The electronics of the cameras of the \hess\ system are divided into two parts:  the front-end
electronics in 16-pixel \textit{drawers} and the back-end elec\-tro\-nics in a crate at the rear of the
camera. Each of the 60 drawers in a camera contains 16 \PMTs\ and associated acquisition and local trigger
electronic cards. The analogue signals are digitised in the drawers and then the digital signals are
sent to the acquisition crate. Data are sent to the central data acquisition system via an optical fibre.
The connection to the array trigger is via an additional optical fibre.

\subsection{The acquisition channels}
\label{electronic}

\begin{figure}[h]
\begin{center}
\mbox{\epsfig{file=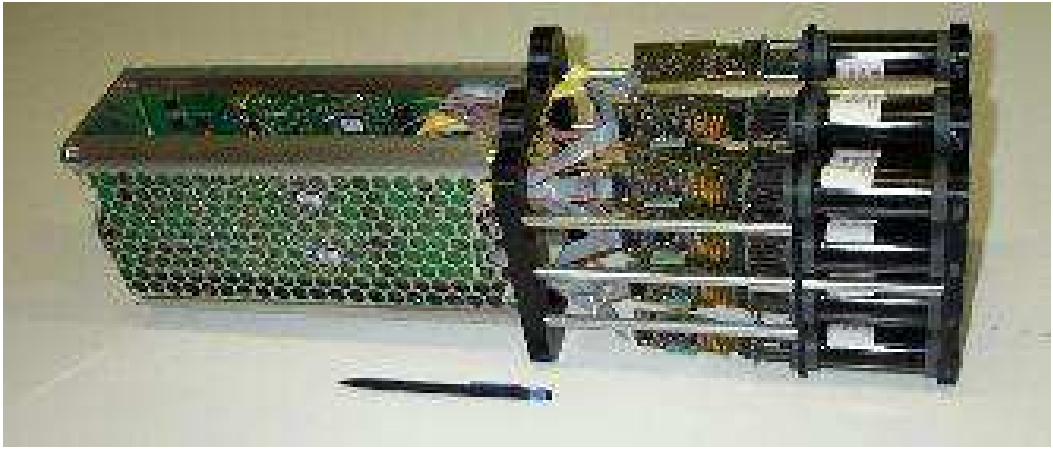,width=1.0\textwidth}}
\end{center}
\caption{A camera drawer with its 16 \PMTs, the individual HV supplies for each \PMT,
the two 8-channel acquisition cards and the control/interface card.}
\label{fig:Drawer}
\end{figure}

\begin{figure}[h]
\begin{center}
\mbox{\epsfig{file=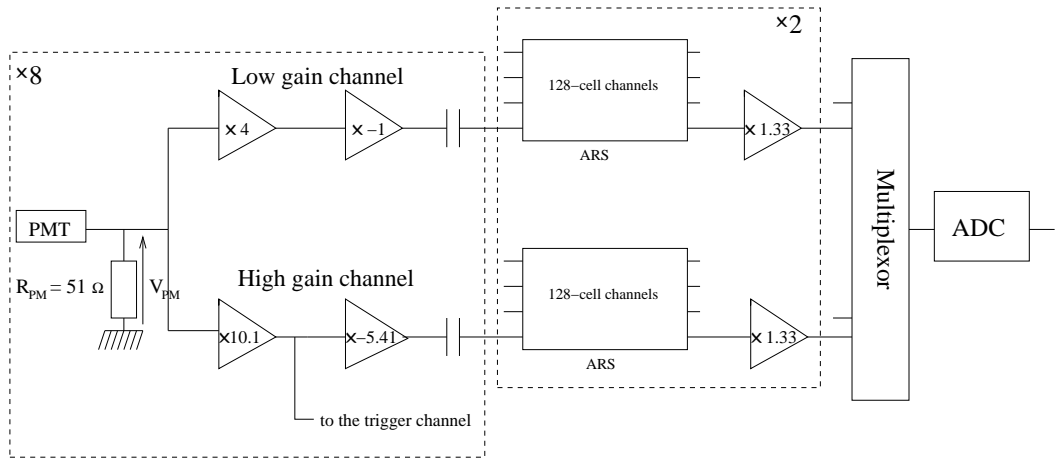,width=1.0\textwidth}}
\end{center}
\caption{Schematic illustration of the electronics of one acquisition card ($8$ \PMTs).}
\label{fig:AcquisitionChannels}
\end{figure}

\begin{figure}[h]
\begin{center}
\mbox{\epsfig{file=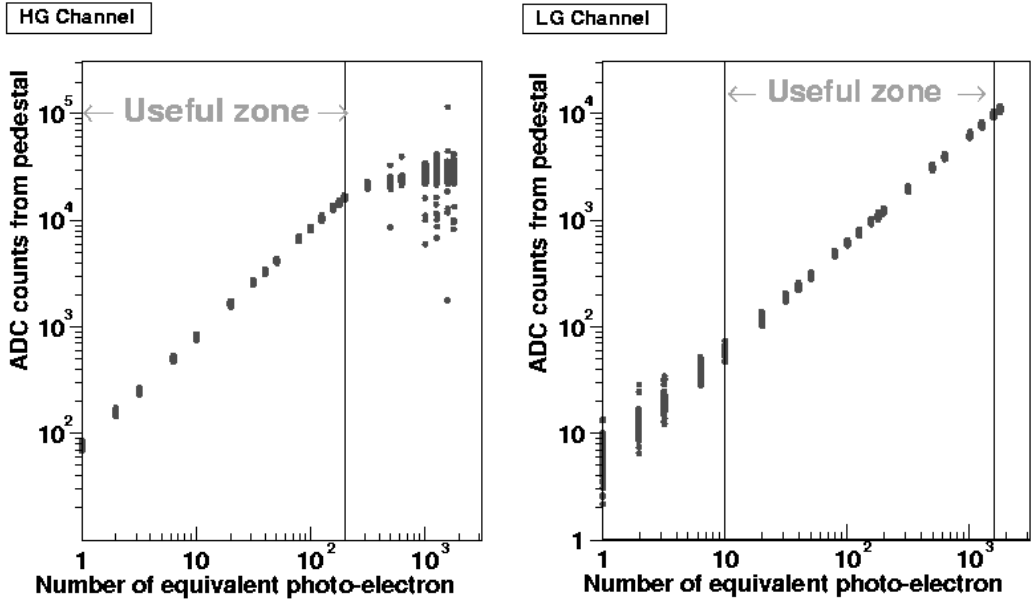,width=1.0\textwidth}}
\end{center}
\caption{Linearity of the high gain channel (left-hand side) and low gain channel (right-hand side). 
Measured signal in \ADC\ counts is shown against input signal in photoelectrons. 
The useful zones used for the analysis are shown.}
\label{fig:DynamicRange}
\end{figure}

Each drawer ($16$ pixels, see Figure~\ref{fig:Drawer}) is composed of two acquisition cards, each reading the
data from $8$ \PMTs, as shown in Figure~\ref{fig:AcquisitionChannels}. For each pixel, there are three channels,
one trigger channel and two acquisition channels with different gains: the high-gain (HG) channel is used to
detect signal charges up to $200$ photoelectrons (p.e.);  the low-gain (LG) channel is used to cover the range
from  $15$ to $1600$ photoelectrons. Figure~\ref{fig:DynamicRange} illustrates the linear range used for analysis
for both channels.  The \PMT\ signal is measured across a resistor $R\mathrm{_{PM}}$ and amplified into the two
acquisition channels (HG and LG). The analogue signal is then sampled in an Analogue Ring Sampler (\ARS)
initially developed by the CEA for the ANTARES experiment \cite{ars}. The sampling is performed at a rate of $1$
GHz; the analogue voltage levels are stored in a ring buffer consisting of $128$ capacitor cells. Following a
trigger signal, the sampling is stopped, the capacitor cells are addressed  one by one, their analogue signals
are impedance matched and a multiplexor distributes the signal from $4$ \ARSs\ ($8$ \PMTs) into one Analogue to
Digital Converter (\ADC) with a conversion factor $V\mathrm{_{ADC}}$ of $1.22$ mV/(\ADC\ count).Only $N_{L}$ (for
normal observations set to 16) cells are  converted into charge equivalent ADC counts, in the  range where
the Cherenkov signal is expected on the basis of the trigger  timing. The Section~\ref{timing} describes how this
timing is calibrated. The digitised signals are stored and processed in a   field-programmable gate array (FPGA)
on each acquisition card.   In the normal read-out mode  (\textit{charge  mode}), the $N_{L}$ samples are summed
to give two \ADC\ values per pixel (HG and LG). Data from the drawers are  sent to 8 FIFO memories located  in a
cPCI (Compact Peripheral Component Interface)  crate to be read back by the  CPU. 

The high-gain channel is sensitive to single photoelectrons: the number of \ADC\ counts between the pedestal
and the signal from a single photoelectron is set to be $\approx 80$ \ADC\ counts (for a \PMT\ gain of
$2\times10^{5}$). This value is chosen such that the single photoelectron peak can be clearly distinguished
at normal pixel operating high voltage (HV).  The electronic noise of the system is such that the pedestal
width is typically 20\% of a photoelectron ($\approx 16$ \ADC\ counts \RMS). The low electronic noise and good
sampling allow a precise gain calibration of every pixel.

\subsection{\PMT\ high voltage supply and slow control}

In addition to the acquisition cards, a drawer contains a number of monitoring functions including three
temperatures sensors, as well as individual high voltage supplies for each channel, which are controlled by
the control/interface board. The high voltage supplies are DC-DC converters with active regulation for the
cathode  (on negative high voltage) and the four last dynodes. The high voltage is set via an analogue level
supplied by the control board, and is adjusted to provide a uniform \PMT\ gain across the camera. In order to
minimise transit time differences between \PMTs\ caused by the different high voltages applied, \PMTs\ were
sorted into drawers and cameras according to their gain. A high voltage divider and \ADC\ allows the actual
high voltage to be read back. In order to monitor pixel performance and to identify pixels receiving light
from stars, both the total current provided by the high voltage supply to the \PMT\ and the divider chain
(``HVI'') as well as the DC anode current (``DCI'') are monitored. The ``DCI'' current depends on the \PMT\ 
illumination and the electronic chain offset.

\subsection{Calibration parameters}
\label{amplitude}

Standard Cherenkov analysis use as a starting point the signal amplitude of each pixel. The amplitude is the
charge in p.e. induced by light on the \PMT, this charge being corrected by the relative efficiency   of this
pixel compared to the mean value over the camera (``flat-fielding'', see Section~\ref{ff}).  The calibration
provides the required conversion coefficients from \ADC\ counts to  corrected photoelectrons.

For each event, \ADC\ counts are measured in both channels, $ADC^{HG}$ for the
high gain channel and $ADC^{LG}$ for the low gain channel. The calculation of the
amplitude in photoelectrons received by every pixel is, for both
channels:
\begin{eqnarray}
\qquad \qquad A^{HG} & = & \frac{ADC^{HG}-P^{HG}}{\gamma_{e}^{ADC}} \times FF \nonumber \\
\qquad \qquad A^{LG} & = & \frac{ADC^{LG}-P^{LG}}{\gamma_{e}^{ADC}} \times (HG/LG) \times FF
\end{eqnarray}
where,
\begin{itemize}
\item $P^{HG}$ and $P^{LG}$ are the \ADC\ position of the base-line for both channels,
which are the \textit{pedestal} positions,
\item $\gamma_{e}^{ADC}$ is the gain of the high gain channel in \ADC\ counts per p.e., 
\item $HG/LG$ is the amplification ratio of the high gain to the low gain,
\item and $FF$ is the flat-field coefficient.
\end{itemize}
The flat-field coefficient corrects for different optical and quantum 
efficiencies between pixels within a camera.

For the image analysis, $A^{HG}$ and $A^{LG}$ are used to provide a single pixel amplitude.
Provided that both gain channels function properly, the HG value alone is used up to $\approx 150$~p.e.  (see
Figure~\ref{fig:DynamicRange}) and the LG value beyond $\approx 200$~p.e. For intermediate values, a weighted
average of the HG and  LG values is used and the amplitude is given by:
\begin{eqnarray}
\qquad \qquad A  = (1-\epsilon) \times ADC^{HG} + \epsilon \times ADC^{LG} 
\end{eqnarray}
where $\epsilon \approx (ADC^{HG}-150) / (200-150)$.

To conclude, for both channels of every pixel, the calibration must provide  the pedestal positions,
the high gain $\gamma_{e}^{ADC}$, and the ratio of the high gain to the low gain.  As the
flat-fielding coefficients do not depend on the electronics, they are calculated per pixel and not
per (HG or LG) channel as will be seen later (see Section~\ref{ff}). It is also essential that any
non-operational channels or pixels are identified in the calibration process in order to avoid
miscalculation of amplitudes.

\section{Timing of the readout window}
\label{timing}

A time window containing only $N_L$ cells (from the 128 available) is read from the \ARS\ (see Section
\ref{electronic}). These cells are selected as follow. When the reading command is received, the
\ARS\ stops sampling and the first cell of the readout window is the $(128-N_d)^{th}$ sample in the
past from the latest filled sample in the circular buffer. $N_d$ is the time between the moment the
signal is received by the pixel and the moment the trigger signal comes back to the drawer. The $N_d$
value has then to be calibrated in order that the Cherenkov signal integrated over the $N_L$ cells is
maximised.

The position of the readout window (the $N_d$ value) is verified frequently using the {\it sample
mode} facility of the \ARSs. In this mode, the charge of the $N_L$ cells (each of 1~ns) is read and
stored. The pulse shape can be then studied, as well as the readout timing. The position of the charge
peak in the readout window is used to adjust the timing.

\begin{figure}[h!]
\begin{center}
\psfig{figure=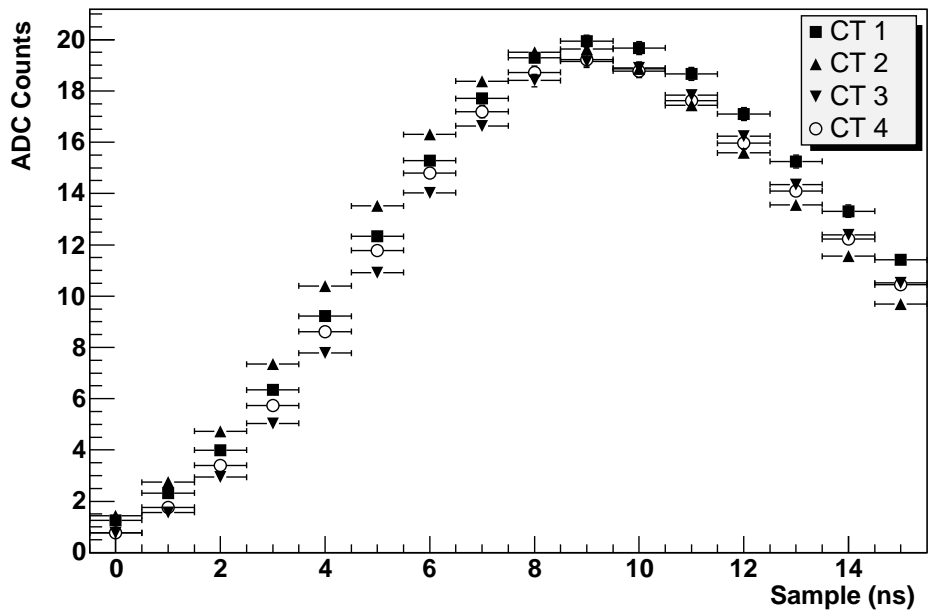,width=1.0\textwidth}
\end{center}
\caption[]{Mean signal samples over HG channels for the four cameras for air-shower events.}
\label{fig:sample}
\end{figure}

Figure~\ref{fig:sample} gives an example of data from a {\it sample mode} run made with air-shower
events. The electronic noise in each cell, obtained using events which do not contain contributions due
to Cherenkov light, is subtracted to extract the pulse shape. They are averaged over the HG channel for
each telescope. From the \ARS\ sample mode, the accuracy of timing of the readout window is about 1~ns.

\section{Estimation of pedestals}
\label{pedestal}

The pedestal position is defined as the mean \ADC\ value recorded in the absence of any Cherenkov light.

In the dark, electronic noise creates a narrow Gaussian \ADC\ distribution whose mean is the pedestal
position. However, the pedestal position has some dependence on camera temperature, which varies
between $20^\circ$C and $40^\circ$C depending on the season and the weather. The typical temperature variation
during a run is of the order of $1^\circ$C.
The pedestals must therefore be calculated for all channels during time-slices
as short as possible to minimise the impact of pedestal drift due to temperature variations in the
electronics.

During observations, the pixels are illuminated by night-sky background (\NSB) photons, greatly  increasing the
pedestal width. There is over 1~p.e. of \NSB\ per readout window in normal operation. Due to the capacitive
coupling of the \PMT\ signals to the \ARS\ (see Figure~\ref{fig:AcquisitionChannels}) the pedestal position
remains constant in the usual  \NSB\ range in Namibia. The mean pedestal position therefore corresponds to the
position in \ADC\ counts of zero photoelectrons of Cherenkov light.

\subsection{Dark pedestal}

To measure precisely the electronic noise in all channels, \ADC\ distributions are taken in the absence
of background light. The lids of all cameras are closed and the \ADC\ distributions are randomly measured by
software triggering of the cameras when the internal camera temperature is stable.  

\begin{figure}[h!]
\begin{center}
\psfig{figure=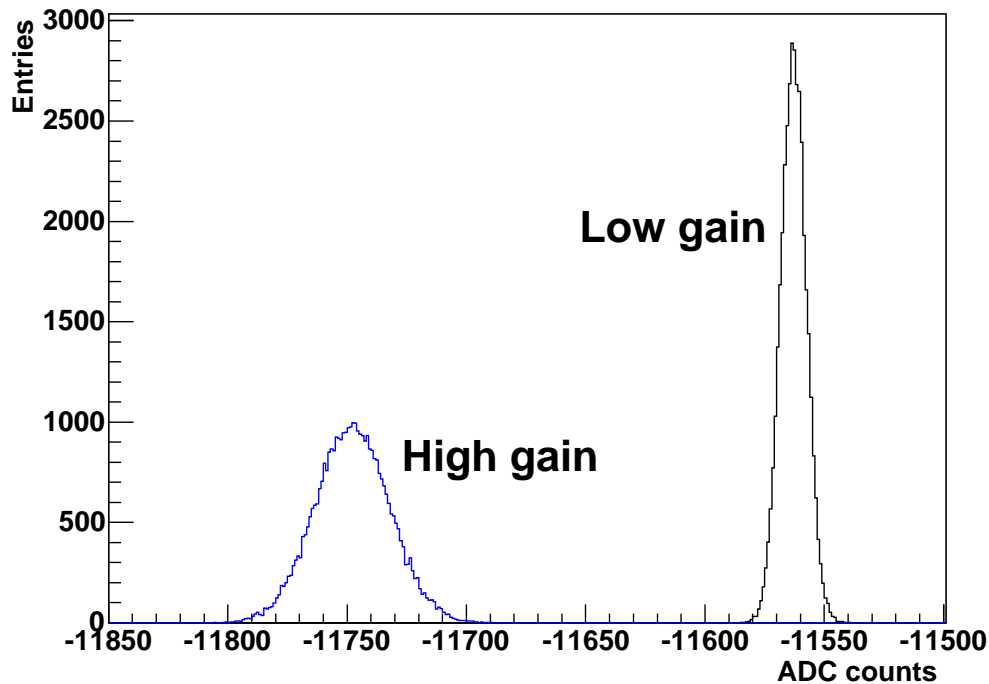,width=1.0\textwidth}
\end{center}
\caption[]{\ADC\ distributions of dark pedestals for the high and low gain 
channels of a single pixel.}
\label{fig:ElecPed}
\end{figure}

The resulting pedestal is determined by the base-line voltage of the electronic channels at the input
of the \ADC. These base-lines are approximately  $-0.9$ V for both channels, corresponding to $\approx
-730$ \ADC\ counts.  In \textit{charge mode}, after the summation of the $16$ samples, the \ADC\ counts are
then $\approx -11500$ \ADC\ counts: this is the typical electronic pedestal position. Measured pedestals
lie in the range from -13000 to -11000 \ADC~counts.  Random noise from \PMTs\ and from electronic
components is responsible for the pedestal width (see Figure~\ref{fig:ElecPed}).  The noise at the 
input of the \ADC\ is about $20\,\mathrm{mV}$ in the high-gain channel, which gives a pedestal \RMS\ of
$16$ \ADC\ counts or about 0.2 p.e., and $7\,\mathrm{mV}$ in the low-gain channel, which gives a
pedestal \RMS\ of $6$ \ADC\ counts or about 1 p.e.\\

Runs taken with the camera lids closed can also be used to measure the drift of pedestal values with temperature using
the temperature sensors mounted at three positions in every drawer. The dark pedestal position is correlated with
temperature. The overall behaviour is a decreasing pedestal position for increasing temperature. As the shift can be as
large as -50 \ADC~counts/degree, pedestal positions for observation runs must be calculated frequently (roughly
every minute) to achieve the required precision ($\ll$1~p.e.).

\subsection{Pedestal with night-sky background}

During an observation run, the \NSB\ modifies the \ADC\ distributions of pedes\-tals. As the coupling
between \PMT\ and \ARS\ behaves like a RC circuit, the short photoelectron pulses (of positive polarity
due to the inverting amplifiers)  are followed by a slight negative undershoot over a few
micro-seconds. For typical \NSB\ rates, of order 100~MHz (in units of p.e. per second), the time
between \NSB\ photoelectrons is short compared to the time constant of the undershoot. The undershoots
combine and average, causing a negative shift of the base-line, onto which \NSB\ photoelectron signals
are superimposed in a way that the overall average level remains at the dark-pedestal value.

\begin{figure}[h!]
\begin{center}
\psfig{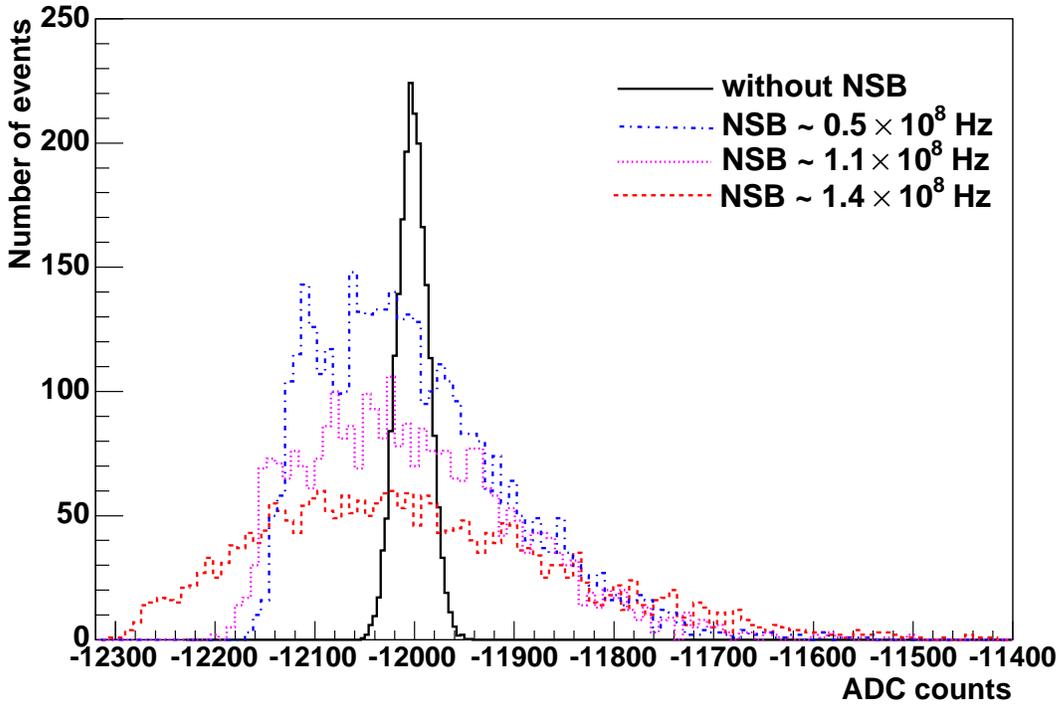}
\end{center}
\caption[]{\ADC\ distributions of pedestals at different night-sky background
rates in real data.}
\label{fig:NSBPed}
\end{figure}

When pedestals are measured during observation runs, the resulting distributions depend on the level of
the \NSB. For low \NSB, it can happen that no \NSB\ photoelectron arrives within the 16~ns integration
window (see Section~\ref{electronic}); in this case the measured pedestals have a similar narrow width
to the dark pedestals, but  exhibit a negative shift. Combining such events with events with one or
more \NSB\ photoelectrons which are  fully or partly contained in the integration window,  one finds a
pedestal distribution with a sharp rise and peak at the location of the shifted dark pedestal, followed
by  a smeared single-photoelectron peak and a tail towards higher values (distribution for 50~MHz \NSB\ 
in Figure~\ref{fig:NSBPed}).

At higher \NSB\ levels, well above 100~MHz, there are usually several photoelectrons within the
integration window and the pedestal distribution is essentially Gaussian, with a width given by the
square root of the mean number of \NSB\ photoelectrons (up to small corrections for the electronic
noise, the width of the single-photoelectron amplitude distribution and the effect of photoelectron
signals truncated by the integration window). The two other distributions shown in
Figure~\ref{fig:NSBPed} for 110~MHz and 140~MHz \NSB\ rates represent intermediate cases, where the
asymmetric shape of the Poisson distribution in the number of  photoelectrons in the integration
window is still visible.\\

Pedestals for observation runs are determined as often as possible in order to account for camera temperature
variations. As a shower image contains typically only 20~pixels, real triggered events are used to measure
pedestals. Pixels containing Cherenkov light are rejected. To identify such image pixels, amplitudes (in
p.e.) are roughly calculated using the previously-calculated  pedestal and using the nominal gain of channels
(80~\ADC\ counts per p.e. for the HG and 6 for the LG). For a given pixel, if neighbouring pixels have a signal
amplitude above a threshold (around 1.5 to 3~p.e.) or if its own amplitude is above a threshold, this pixel
is suspected to be contaminated by Cherenkov light.  So this pixel's value for this event does not contribute
to this pixel pedestal histogram. This algorithm is applied for all pixels and all events until there are
enough entries in pedestal histograms to determine accurately the pedestal positions from the mean of the
histogram. The frequency of pedestal updates depends on the event rate of the observation run; pedestals are
calculated approximately every minute.

\section{Estimation of night-sky background}
\label{nsb}

Knowledge of the actual \NSB\ rate in each pixel -- possibly influenced by a star in the  field of view
of the pixel -- is important to judge if a pixel can be included in the image analysis. The \NSB\ rate
can be determined in two ways,  from the pedestal width as illustrated in the previous section, and
from the \PMT\ currents.

\subsection{\NSB\ from pedestals width}

One method to estimate the \NSB\ brightness is to use the pixel pedestal widths. This technique is used as
standard by the VERITAS collaboration \cite{bond}. During an observation,  the pedestal width is dominated by
\NSB\ photon fluctuations. The typical pedestal \RMS\ is about 100~\ADC\ counts for the HG channel, corresponding
to roughly 1.2 p.e.  The \NSB\ rate is estimated from the square of the pedestal width  (subtracting the
electronic noise and the charge dispersion of a single photoelectron in quadrature): 
\begin{eqnarray}
\label{eqn:pednsb}
F_{NSB} = \sqrt{RMS_{P}^2-RMS_{0}^2-\sigma_{\gamma_{e}}^2}\, /\, (N_{L} \, \tau) \,\,\mathrm{Hz}
\end{eqnarray} 

where $RMS_{P}$ is the pedestal \RMS\ in p.e., $RMS_{0}$  is the electronic pedestal \RMS\ in p.e.,
$\sigma_{\gamma_{e}}$ is the charge dispersion in p.e. of a single photoelectron and  $ N_{L} \times \tau
\,=\, 16\times10^{-9}$ is the readout window in seconds. This expression should be accurate within 10-20\%
considering the corrections mentioned earlier.

\subsection{\NSB\ from \PMTs\ currents}

\begin{figure}[h!]
\begin{center}
\psfig{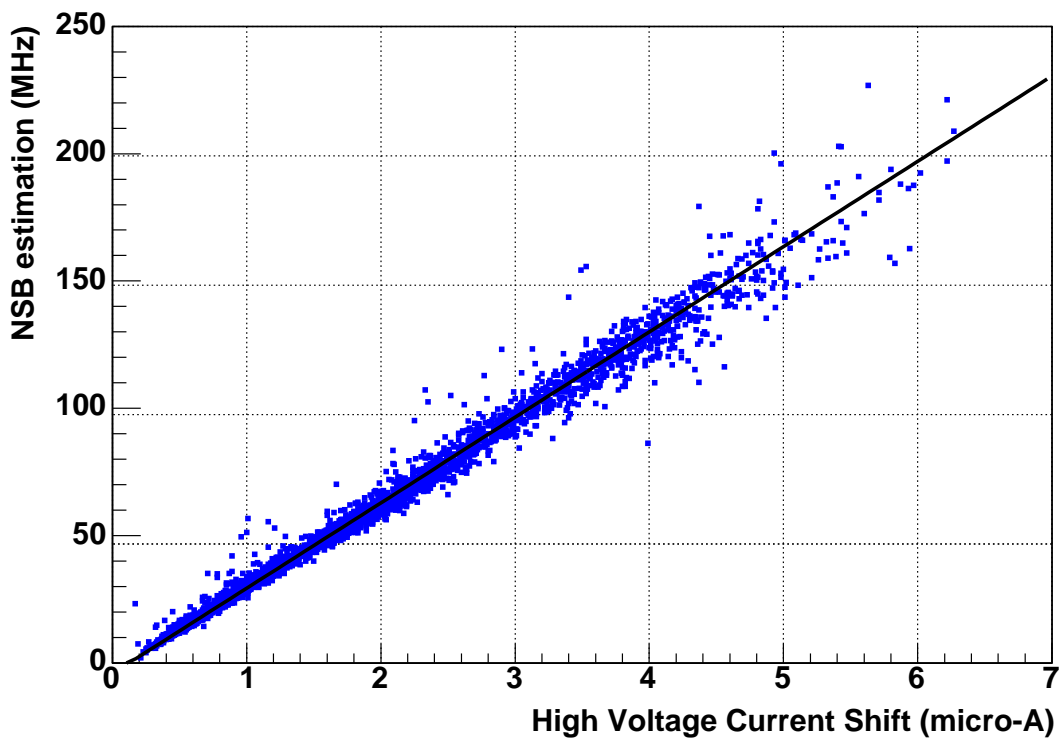}
\end{center}
\caption[]{Offset-subtracted HVI current
 as a function of \NSB\ brightness for all the pixels of one camera, measured in a test setup with controlled
 illumination.}
\label{fig:nsb}
\end{figure}

Because of its small temperature dependence and its larger range, the total current drawn by each \PMT\ from the
HV supply (HVI) is used to estimate the \NSB, rather than the anode current (DCI). The HVI current is composed
of the current drawn in the  absence of illumination (mainly the divider current, proportional to the high
voltage)  and a component which increases linearly with the anode current. The shift of the HVI current from
its dark value against DC p.e. flux ($F_{NSB}$) is shown in Figure~\ref{fig:nsb}. The relationship is 
described by Equation~\ref{eqn:hvinsb}.

\begin{eqnarray}
\label{eqn:hvinsb}
F_{NSB} = (3.235\pm0.006) \times 10^{7} \times \Delta HVI_{\mu A}
\end{eqnarray} 

To estimate the \NSB\ with this method, one needs to know the values of the HVI in the dark. Values of the
HVI base-lines, $\mathrm{HVI_{0}}$, are extracted regularly from runs taken with the camera lids closed. 
The \ADC\ used to measure HVI has a very small temperature dependence (with slopes in the range -0.02 to
+0.03~$\mathrm{\mu A/}$\dg C) compared to the current produced by normal values of \NSB, $10^{2}$~MHz,
which is of the  order of 3.2$~\mathrm{\mu A}$. This temperature dependence can be neglected in the estimation
of \NSB.\\

\begin{figure}[h!]
\begin{center}
\psfig{figure=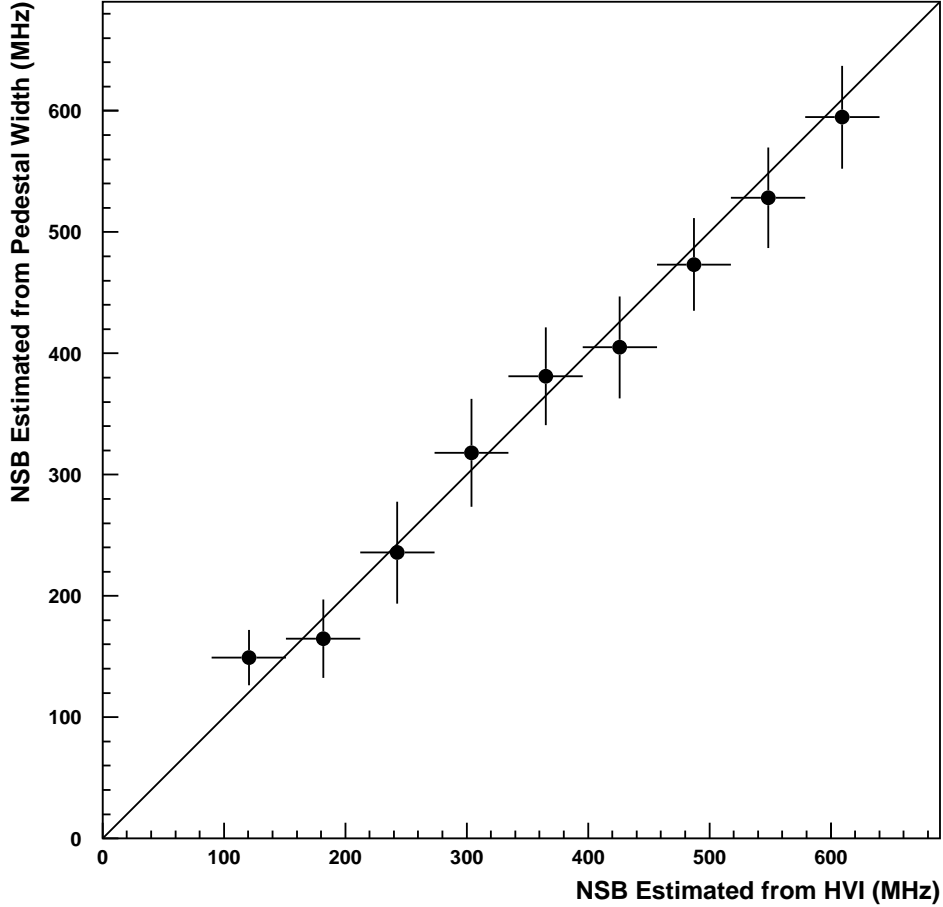,width=1.0\textwidth}
\end{center}
\caption[]{Correlation of \NSB\ estimates using HVI and pedestal width. The y-error bars are the \RMS\ of the 
y-distribution. The line drawn corresponds to a perfect correlation.}
\label{fig:nsbped}
\end{figure}

The correlation between the two estimates of the \NSB\ is given in Figure~\ref{fig:nsbped} for one camera  for
a single observation run, and are in reasonable agreement. A \NSB\ rate of 250~MHz corresponds approximately to
the presence of a star of magnitude $\sim4$.  Pixels with noise levels above this can be treated differently
in the Cherenkov image analysis and are flagged as noisy.

\section{Conversion factor between \ADC\ counts and signal charge}
\label{gain}

The \PMT\ signal, $V_{PM}$ (see Figure~\ref{fig:AcquisitionChannels}),  is measured across a resistor and
is amplified into two acquisition channels, one with low gain $G^{LG}$ and the other with high gain $G^{HG}$. The
amplified signals are then converted into a digital signal with $\tau = 1$~ns samples and summed over
16~ns. The conversion factor of the \ADC\ is approximately $V_{ADC}$=1.22mV/Count. The number of 
(summed) \ADC\ counts   equivalent to one photoelectron is (for both readout channels $i$): 
\begin{eqnarray} 
\qquad \qquad \qquad \gamma_{e,i}^{ADC} = \frac{G^{i}}{\tau \, V_{ADC}} \, 
\int V_{PM,s.p.e.}(t)~\mbox{d}t 
\end{eqnarray} 
Here $V_{PM,s.p.e.}$ is the single photoelectron pulse shape. We define the pixel \textit{gain} as
this conversion factor. This factor includes the \PMT\ gain, the signal amplification in both readout
channels and the integration in the \ARS. 

\subsection{LED pulser for single-photoelectron illumination}

Special runs are taken roughly every two days to determine the pixel gains. In these runs a LED pulser is
used to illuminate the camera with an intensity of about 1 photoelectron/pixel. The LED and a diffuser are
installed two meters in front of the cameras in their shelter. The light intensity homogeneity over the
camera is $50$~\%. The calibration LED is  pulsed at 70~Hz and the same signal which pulses this LED is  used
to trigger the camera acquisition (after a suitable fixed delay) such that the  signal arrives centered in
the readout window (as explained in Section~\ref{timing}).

\subsection{Determination of the gain of the HG channel}
\label{HG}

The pixel gain can be extracted from data taken under $\sim 1$ photoelectron illumination via a fit to
the \ADC\ count distribution of each pixel.  The \ADC\ distribution fit function is derived from the
following assumptions: the number of photoelectrons follows a Poisson distribution, the electronic noise
is much smaller than the  width of the single~p.e. distribution, and the 1~p.e. distribution is described
by a Gaussian distribution. Then, the electronic pedestal is approximated by a Gaussian with a
standard deviation $\sigma_P$ and with a mean position in \ADC\ counts $P$. The light distribution for a
given signal of $n$~p.e. ($n\in\mathbb{N}$) is approximated also by a Gaussian with a standard
deviation $\sqrt{n}~\sigma_{\gamma_{e}}$ and with a mean position in \ADC\ counts
$P+n\,\gamma_{e}^{ADC}$,  $\gamma_{e}^{ADC}$ being the conversion factor between \ADC\ counts and
photoelectrons and $\sigma_{\gamma_{e}}$ is the \RMS\ of the charge induced by a single photoelectron. 
Under a mean light intensity $\mu$ the expected signal distribution is given by:

{\setlength\arraycolsep{2pt}
\begin{eqnarray}
\label{eqn:SinglePe}
\mathcal{G}(x) = N\times & \bigg( & \frac{e^{-\mu}}{\sqrt{2\pi}\sigma_P} 
   \, exp\bigg[ -\frac{1}{2} \Big(\frac{x-P}{\sigma_P}\Big)^2 \bigg] \nonumber \\
 & + & \kappa \sum_{n=1}^{m\gg1} \frac{e^{-\mu}}{\sqrt{2\pi}\,\sigma_{\gamma_{e}}} \frac{\mu^n}{n!}
   \, exp\bigg[ -\frac{1}{2} \Big(\frac{x-(P+n\,\gamma_{e}^{ADC})}{\sqrt{n}\,\sigma_{\gamma_{e}}}\Big)^2 \bigg] \bigg)
\end{eqnarray}}

\begin{figure}[h!]
\begin{center}
\psfig{figure=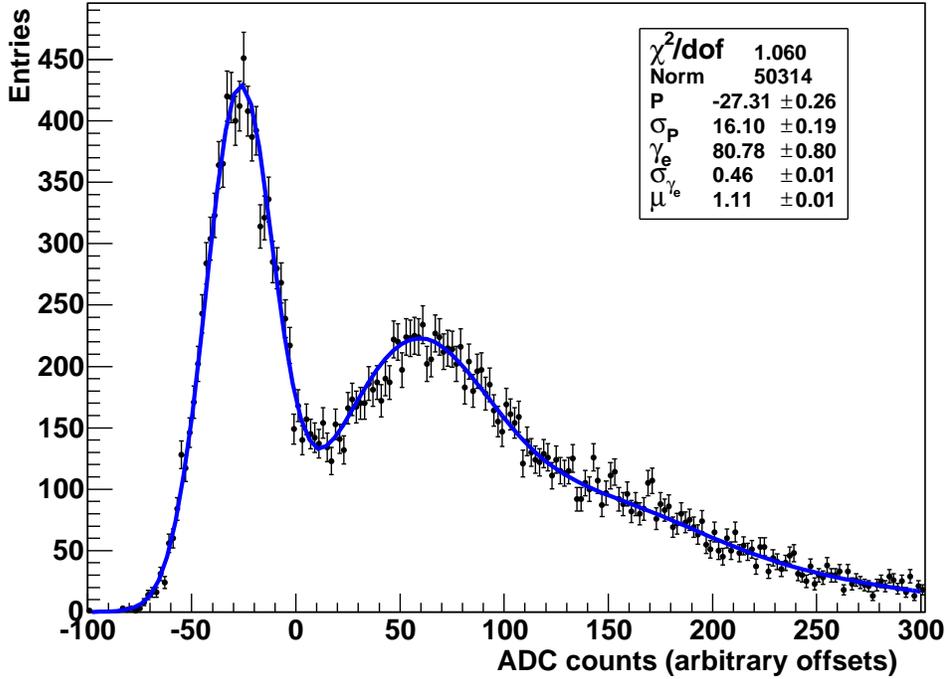,width=1.0\textwidth}
\end{center}
\caption[]{Example of an \ADC\ count distribution for a single-photoelectron run.}
\label{fig:ADCtoPe}
\end{figure}

This function is used to fit the signal distribution of each pixel. All parameters are free except the overall
normalisation $N$ which is fixed according to the number of events in the run. The normalisation $\kappa$ should be equal
to one for a Poissonian statistics and can also be adjusted. A sample signal distribution and fit are shown in
Figure~\ref{fig:ADCtoPe}.

The high-gain values of $\gamma_{e}^{ADC}$ for the four telescopes are given in Table~\ref{tab:calib_cam}. In
January 2004 the mean gains are close to the nominal value of 80~\ADC\ counts per p.e. and
their \RMS\ is around 2-3~\ADC\ counts for all telescopes.  The distributions of $\gamma_{e}^{ADC}$ over pixels 
are almost Gaussian and the pixel gains are randomly distributed within the cameras. 

\begin{table}[h!]
  \begin{center}
    \begin{tabular}{|c||c|c|c|c|c|}
      \hline
\multicolumn{1}{|c||}{Telescope}  &\multicolumn{2}{c|}{$\gamma_{e}^{ADC}$}     
                    &\multicolumn{2}{c|}{$HG/LG$}    &$FF$\\

        &mean &\RMS\  &mean  &\RMS\   &\RMS\  \\ \hline
CT 1    &79.8 &1.8  &12.98 &0.44  &0.10 \\ \hline
CT 2    &79.4 &3.0  &13.73 &0.43  &0.11 \\ \hline
CT 3    &76.8 &3.5  &13.32 &0.47  &0.13 \\ \hline
CT 4    &77.3 &3.8  &13.52 &0.63  &0.09 \\ \hline
      \end{tabular}
    \vspace{3mm}
    \caption{Summary of mean values of calibration coefficients in January 2004,
    and of their variation across pixels in a camera. 
    The high-gain $\gamma_{e}^{ADC}$ values are given in \ADC\ counts.
    By definition, the mean flat-field coefficient is 1.}
    \label{tab:calib_cam}
  \end{center}
\end{table}

However, the fit may fail when the light intensity $\mu$ of the $\gamma_{e}^{ADC}$ measurement is too low
or too high or when the corresponding \ARS\ does not work properly (see Section~\ref{bp}). No gain
estimate is then available for this pixel and a gain of 80 \ADC\ counts/p.e. is assumed. In this case, the
flat-field coefficient compensates for any deviation from the  nominal gain (see Section~\ref{ff}).

Laboratory measurements show that the single photoelectron distribution of the Photonis~XP2960 \PMTs\ used by
\hess\ is significantly non-Gaussian. The fit to the signal distribution assuming a Gaussian shape leads to a
bias of a few percent in the determined gain. This bias is included and accounted in our  simulations (which
use the measured single-p.e. distribution). 

\subsection{Determination of the gain of the LG channel}

In the regime where both gain channels are linear (30 to 150~p.e. in order to be conservative - see
Figure~\ref{fig:DynamicRange}) the ratio of the signals provides a measurement of the amplification ratio in
the two channels.  With the previously calculated gain for the HG channel and with this amplification ratio,
the gain of the LG channel can be calculated. 

Cherenkov events from normal observations are used to measure these gain ratios. Using nominal gains,
approximate pixel charges are determined and pixels with charges in the overlap regime are selected. 
The mean ratio of the pedestal-subtracted \ADC\ counts in both channels for these selected events is
used to estimate the amplification ratio for each pixel.

The gain ratios for all cameras are given in Table~\ref{tab:calib_cam}. The mean ratios are around
$13.5$ (with \RMS\ $\sim 0.5$). The slight differences are due to the gain dispersion of the amplifiers.

\section{Flat-field coefficients}
\label{ff}

Different photocathode efficiencies  and different Winston cone light
collection efficiencies \cite{optics1} produce some inhomogeneity in the camera. Thus, even after 
the calibration of the electronic chain, channels exhibit a slightly  different response to 
a uniform illumination. Flat-field coefficients (\textit{FF}) are used  to correct for 
these differences. 

The flat-field coefficients are determined from special flat-field runs (taken roughly every 2 days)
which use LED flashers~\cite{durham2} mounted on the telescope dishes at 15~m from the cameras 
to provide uniform illumination.
The flashers produce short pulses (FWHM of 5~ns) with uniform illumination out to 10\dg\ (bigger than
the angular size of the camera). The wavelength range of these pulses is 390 to 420~nm, around the
wavelength of \PMT\ peak quantum efficiency. The pulse intensity is stable within 5\% \RMS\ and can be
adjusted by the use of 5 different neutral density filters, giving an operating range of 10 to 200~p.e.
The camera is triggered in the same way as for Cherenkov showers but with an increased pixel
multiplicity ($>$9 pixels) to reduce the  background of air-showers.

Flat-field coefficients are extracted from flat-field run data using calibrated amplitudes without the FF
correction. For each event the mean signal over each camera is calculated, excluding unusable pixels
(determined as described in the Section~\ref{bp}). The ratio of each pixel signal to this mean signal is
accumulated in a histogram. The mean of this  ratio over the run gives the efficiency of each pixel relative to
the camera mean. The inverse of this mean value, the \textit{flat-field coefficient} (\textit{FF}), is used to 
correct for pixel efficiency differences. By definition, the mean \textit{FF}  is equal to 1. The distribution
of flat-field coefficients gives an estimate of the  uniformity of each camera. As can be seen from
Table~\ref{tab:calib_cam}, the  typical \RMS\ of the \textit{FF} distribution is $\approx 10$\%.

\section{Stability of calibration parameters}
\label{stability}

\subsection{Stability within a lunar cycle}

It is expected that the calibration coefficients are stable over periods of weeks if the \PMT\ high
voltage is not changed. During each observation period (corresponding  to a lunar cycle) several
measurements of each calibration value are made for every pixel. The \RMS\ variation of these
measurements is used to characterise the stability of the calibration parameters. The values of the \RMS\ 
variation of parameters, averaged over all channels of a camera, are summarised in
Table~\ref{tab:stability} for the January 2004 period  (during which no HV adjustments were made). As
expected, all calibration parameters were stable at the few percent level during this period for all
cameras.

\begin{table}[!h]
  \begin{center}
    \begin{tabular}{|r||c|c|c|c|} \hline
  Telescope                 &CT 1   &CT 2     &CT 3    &CT 4    \\ \hline
  $\gamma_{e}^{ADC}$: average \RMS\ variation &2.1\%  &3.4\%    &3.6\%   &3.4\%   \\
  $HG/LG$: average \RMS\ variation           &1.4\%  &1.2\%    &1.2\%   &1.2\%   \\
  $FF$: average \RMS\ variation               &0.8\%  &0.9\%    &0.8\%   &1.0\%   \\
\hline
      \end{tabular}
     \vspace{3mm}
    \caption{Summary of the stability of calibration parameters in the January moon cycle of 2004. 
     Shown is the \RMS\ variation in time of the different parameters; the \RMS\ values are
     averaged over all channels of a camera.}
    \label{tab:stability}
  \end{center}
\end{table}

The majority of calibration coefficients  are affected by changes to the \PMT\ HV. The \PMT\ gains are
proportional to  $V^{\alpha}$, with $\alpha \approx 5.2$. The flat-field coefficients are also a
function of HV via the collection efficiency of \PMTs\ (collection efficiency of electrons between the
photo-cathode and the first dynode). Only the gain ratio of the two electronics channels is independent
of HV. A complete re-calibration is therefore required when HV adjustments are made (roughly every 3-6
months).

\subsection{Merging of calibration parameters}

Given the stability of calibration parameters on the time-scale of weeks it is reasonable to merge values
taken during stable periods to improve the accuracy of the calibration parameters. The reasons for this
merging are as follows. In a single calibration run there are typically a few  pixels which are not well
measured due to one of the problems described in Section~\ref{bp}. The merging also reduces the statistical
error on the final coefficients measurements, and ensures that remaining systematic effects such as
temperature dependencies are much reduced. 

The first step of the merging process is to identify periods with constant HV settings during at
most one lunar cycle. If the \PMT\ HV has been adjusted during a lunar cycle (to compensate for the
decreased gain caused by \PMT\ aging), two calibration periods are created around the HV change.  A
summary of the merged parameter values within each camera is given in Table~\ref{tab:calib_cam} for
the January 2004 calibration period.

Several precautions are taken to ensure the reliability of the merged values. Pixels which experience
problems within calibration runs are identified using the techniques described in Section~\ref{bp}.
In addition calibration runs with a temperature drift of more than 0.5\dg\ C are excluded from the merging to
guard  against biases introduced by \ADC\ pedestal drifts.

\subsection{Long term variations of calibration parameters}

\begin{figure}[h!]
\begin{center}
\psfig{figure=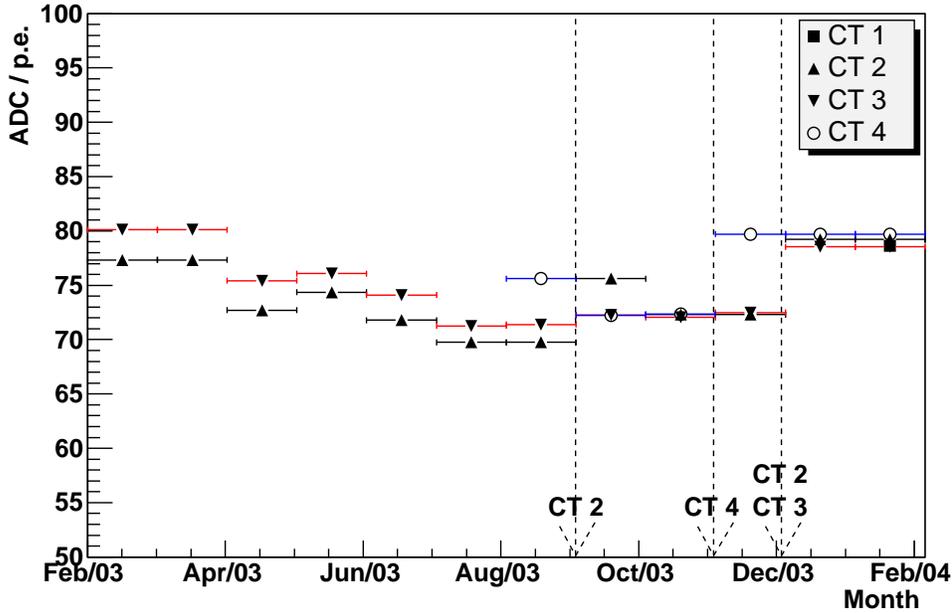,width=1.0\textwidth}
\end{center}
\caption[]{Evolution of the mean $\gamma_{e}^{ADC}$ per camera in 2003/04.
The large arrows correspond to an HV increase for the labelled telescope. 
Telescopes were installed in the following order: CT3, CT2, CT4 and CT1.}
\label{fig:ADCtoPe_evol}
\end{figure}

On time-scales longer than one month, the effects of \PMT\ aging and possibly mirror degradation may
become apparent. Figure~\ref{fig:ADCtoPe_evol} shows the evolution of \PMT\ gain over 1 year. The mean
gain of each camera is shown for each dark moon period.  The general trend is for gains to decrease
with time until a HV adjustment is made. The high voltages were increased for CT2 in October and
December 2003,  for CT4 in November and for CT3 in December.

\begin{figure}[h!]
\begin{center}
\psfig{figure=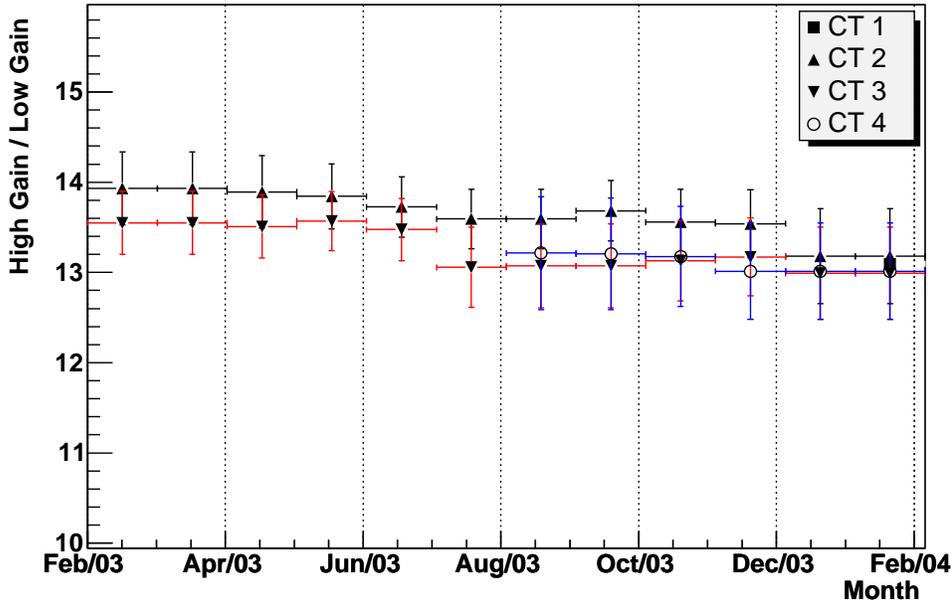,width=1.0\textwidth}
\end{center}
\caption[]{Evolution of the mean HG/LG ratio per camera in 2003/04.}
\label{fig:HLR_evol}
\end{figure}

The HG/LG ratios do not depend on \PMT\ HV and are expected to be stable. The time evolution of the mean
ratio in each camera is given in Figure~\ref{fig:HLR_evol}. 

After December 2003, all drawers were  swapped between cameras to further improve the HV homogeneity within
each camera.  As a consequence, the HG channel gains and gain ratios of January 2004 are not directly
comparable with earlier periods.

\section{Identification of unusable channels}
\label{bp}

In any given run, there are typically a few pixels with characteristics that make  them unsuitable for
use in Cherenkov image analysis. In such cases it may be possible to use the other gain channel to
calculate the signal amplitude (depending on its value).

\subsection{Missing calibration coefficients}

As described in section~\ref{stability}, pixel calibration coefficients are  merged (averaged) for
periods in which the pixel gains are considered stable. Even after this merging process a small number of
pixels do not have useful calibration coefficients. The most common reason for this is that the pixel  is
broken or seriously damaged at the hardware level. The affected gain channels  of such pixels are
excluded from the analysis. 

\subsection{Failure to synchronise the analogue memory}

When an \ARS\ works properly, the signals of the four associated channels are centred in the readout
window of 16~ns. The integrated signal of such channels is at maximum. When an \ARS\ is
\textit{unlocked}, the readout window is misplaced because of timing problems in the window
positioning and the integrated signal is low and unpredictable. As a consequence, the four channels
served by this \ARS\ are not usable. 

This problem appears randomly on the \ARSs\ every time the power is switched on. For a given camera and
gain channel the number of unlocked \ARSs\ ranges from 1-5.  Since the high-gain and low-gain channels of
a pixel are read out by different \ARSs\ (see Figure~\ref{fig:AcquisitionChannels}), it is, however, rare
for both the HG and LG  channels of a single pixel to be unlocked. An estimate of signal amplitude is
therefore usually still available for these channels.

\begin{figure}[h!]
\begin{center}
\psfig{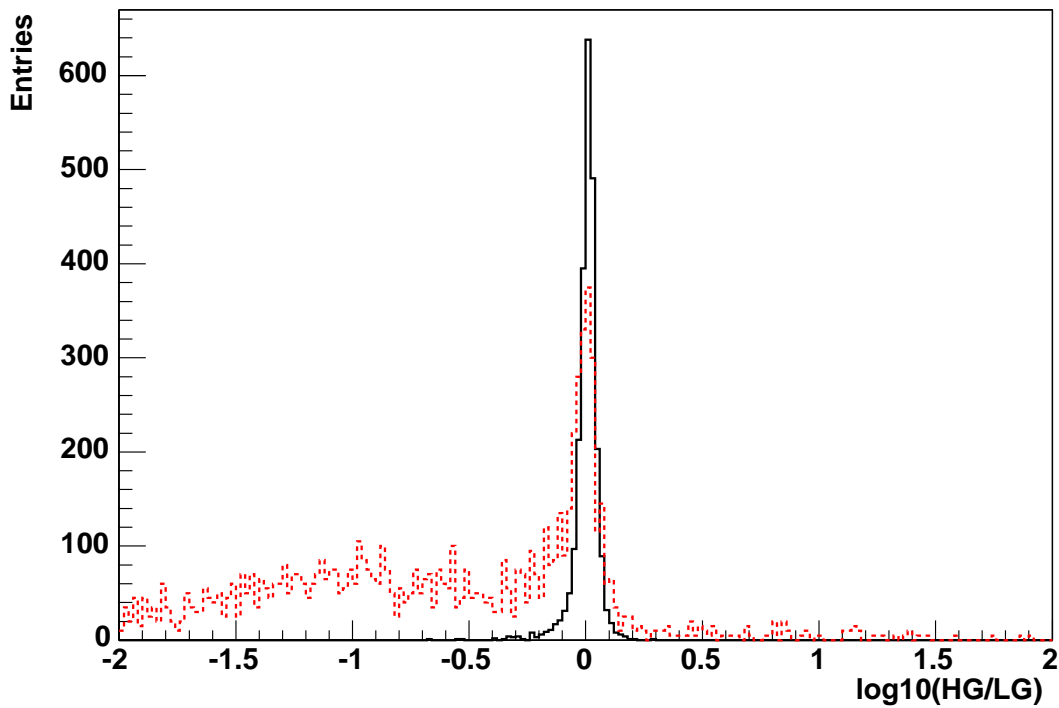}
\end{center} 
\caption[]{Example of pixels on CT2 with a locked (continuous line, \RMS=0.06) and an unlocked \ARS\ in 
the high gain channel (dashed line, \RMS=1.0). The distribution is the logarithm of the ratio of the 
charges calculated from the high and low gain channels in the range 15-150 p.e. The nominal value of 
this parameter is 0.}
\label{fig:BP}
\end{figure}

The distribution of the charge ratio between channels used in the determination  of the amplification
ratio is also used to identify unlocked \ARSs. If one of the gain channels of a pixel is unlocked, the
mean signal of the corresponding channel is less than expected and the signal ratio  distribution is
modified. The \RMS\ of the signal ratio is increased and the mean  value differs from the merged gain
ratio of this period (see Section~\ref{accuracy}). The \RMS\ and the distribution shape are used to flag 
the probably unlocked channels. An \ARS\ is flagged as unlocked if at least two pixels from four have
such amplitude ratio behaviour. Sample amplitude-ratio histograms are presented in Figure~\ref{fig:BP}
for pixels with locked and unlocked \ARSs.

\subsection{Other pixel problems}

The pixel high voltage monitoring information is taken at a rate of
$\sim2$~Hz. Pixels with any deviation during a run of greater than 10~V are excluded from the analysis.

Pixels with a very low frequency of large signals ($>30$ p.e.) are also excluded. Such  pixels may have
a broken \PMT\ base or a damaged photo-cathode. Pixels with a very high frequency of signals above the
same threshold are also excluded, bit errors in the digital  cards of drawers are suspected in this
case.

In addition to pixels with hardware problems there are typically a few pixels in every  field-of-view
that are affected by bright stars. One of the techniques described in  Section~\ref{nsb} is used to
identify these pixels.

These additional hardware problems occur at the $<1$\% level.

\subsection{Summary of channel problems}

Table~\ref{tab:BPstats} summarises the mean number of channels that are considered unusable out of the
1920 electronics channels of each camera in January 2004. This table illustrates that only a small
fraction  of pixels are unusable during observations: less than 2\% of the LG and HG channels have an
unlocked \ARS\ and less than 0.1\% of the channels have other hardware problems. The lack of calibration
coefficients affects less than 2\% of the channels. The fraction of unused channels per run is then of
the order of 3\% to 4\%. The mean fraction of images where a pixel is completely missing due to
hardware problems is 1-2\%.

\begin{table}[h!]
  \begin{center}
    \begin{tabular}{|c||c|c|c|c|}
      \hline
Telescope                   &CT 1 &CT 2 &CT 3 &CT 4      \\
      \hline  
Lack of coefficients        &11   &13   &23   &22   \\
Unlocked HG \ARS\             &6.2  &10.6 &8.2  &13.5 \\
Unlocked LG \ARS\             &15.6 &13.1 &13.6 &18.0 \\
Hardware problem            &0.4  &0.9  &2.3  &0.4  \\
Noisy pixels (Electronics)   &4.4  &0.8  &1.1  &3.0  \\ 
      \hline
      \end{tabular}
     \vspace{3mm}
    \caption{Summary of identified pixels problem. These number are the average number of
    channels with problems (HG or LG) in January 2004. There are 1920 channels in total in 
    each camera.
    }
    \label{tab:BPstats}
  \end{center}
\end{table}

\subsection{Choice of the channel}

The pixel amplitude used in the analysis may be based on the high gain \ADC\ value or the low gain value
or a combination of both, as discussed in Section~\ref{amplitude}. The choice of the channel to be used
for a pixel depends on the amplitude of each channel and the existence of any identified problems with
one of the gain channels. Normally, the HG channel is always preferred for  intensities  below 150~p.e.
and the LG channel above 200~p.e. However, in the case of an unusable HG channel the LG channel is used
down to 15~p.e. In the case of an unusable LG channel, the HG channel is used up to 200 p.e. 

If a pixel has no usable channel, or has a signal amplitude outside of the linear range of its only available
channel, its signal is set to 0~p.e. As mentioned in Section~\ref{HG}, if no high gain ($\gamma_{e}^{ADC}$)
is avaible, flat-fielding coefficient compensate for any deviation from the nominal gain and then the pixel
is usable. But, if no high gain ($\gamma_{e}^{ADC}$) and no flat-fielding coefficient are available, the
pixel is completely excluded. If no gain ratio is available, only the high gain channel is used.

\section{Accuracy of calibration parameters}
\label{accuracy}

\subsection{Pedestal position}

As described in Section~\ref{pedestal}, the \ADC\ pedestal positions depend on temperature. They must
therefore be calculated as frequently as possible to avoid base-line drifts.  To verify that the
calculation frequency is high enough, a method has been developed to  measure the accuracy of the
pedestal positions. 

A typical Cherenkov image involves only $\sim$20 camera pixels. The remaining 98\% of the camera
measures only \NSB\ around the pedestal position. Pixels which are not found in a cluster after 
tail cuts (see \cite{reynolds,bond} for a description of tail cuts technique), are considered to be  
unaffected by Cherenkov light. The mean amplitude of pixels selected in this way  gives an estimate 
of the pedestal position. The precision of the pedestal position can be estimated by comparing this 
value with the previously estimated position.

This method has been applied to an observation run with a large temperature variation (amplitude  of the
order of 1\dg C). The average residual amplitude over the camera is calculated for each new pedestal. This
average is constant in time and is around 0.05~p.e. The residual \RMS\ across the camera is around 0.03~p.e.
We can conclude that measurements of the high gain  pedestal position are frequent enough and are accurate at
the level of 0.05~p.e.\\

Two independent methods have been developed to compute the pedestal positions and calibration parameters. 
Concerning the pedestal evaluation,  the differences consist mainly of diffe\-rent techniques for rejection of
Cherenkov events. A comparison of these two methods provides a second estimate of the pedestal position
accuracy. Table~\ref{tab:pedcomparison} gives the means and \RMS\ of the differences of the pedestal positions
for both channels of each pixel  during one run in January 2004. The differences are given in photoelectrons. 
The two methods agree within 0.1~p.e. in the high gain channel and 0.5~p.e. in the low gain channel.

\begin{table}[!h]
  \begin{center}
    \begin{tabular}{|r||c|c|c|c|}
      \hline
  Telescope        &CT 1   &CT 2     &CT 3    &CT 4 \\
  \hline
  HG: mean         &-0.02  &-0.03    &-0.01   &-0.01 \\
      \RMS\          &0.08   &0.03     &0.06    &0.13  \\
  \hline
  LG: mean         &-0.03  &-0.15    &-0.02   &-0.06 \\
      \RMS\          &0.38   &0.17     &0.49    &0.36  \\
  \hline
      \end{tabular}
     \vspace{3mm}
    \caption{Summary of the differences between two methods to estimate channel pedestal.
             Shown are the mean and \RMS\ of the difference over all channels of a camera, 
	     in units of photoelectrons.}
    \label{tab:pedcomparison}
  \end{center}
\end{table}

\subsection{Comparison of merged coefficients between independent calibrations}

As was done for the pedestals, two independent approaches have been developed to derive the other calibration
coefficients.  The implementation details of the two calibration schemes differ, but the methods described in
this paper are used in both. The differences are the adjustment or not of the normalisation parameter for the
determination of $\gamma_{e}^{ADC}$ and a different event selection based on their amplitudes for the
determination of $FF$ and $HG/LG$ coefficients. Table~\ref{tab:coeffcomparison} gives the mean and \RMS\ of
the differences pixel to pixel of the calibration coefficients for a period in January 2004. The results are
given as a percentage and show good agreement between the different schemes.

The most crucial calibration factor is the conversion factor from  pedestal-subtracted high
gain \ADC\ counts to effective photoelectrons: $\gamma_{e}^{ADC}/FF$. The \RMS\ differences between the
independent schemes for this quantity are $<4$\%. For the ratio of the two gain channels $HG/LG$ the
different schemes agree at a similar level.

\begin{table}[!h]
  \begin{center}
    \begin{tabular}{|r||c|c|c|c|}
      \hline
  Telescope                    &CT 1   &CT 2     &CT 3    &CT 4    \\
  \hline
  $\gamma_{e}^{ADC}$: mean     &1.1\%  &-2.0\%   &-0.5\%  &-1.8\%  \\
                      \RMS\     &2.7\%  &3.0\%    &2.5\%   &2.8\%   \\
  \hline
  Flat-field: mean             &0.3\%  &0.4\%    &-0.2\%  &0.0\%   \\
               \RMS\            &3.0\%  &4.0\%    &5.4\%   &2.9\%   \\
  \hline
  $\gamma_{e}^{ADC}/FF$: mean  &0.8\%  &-2.4\%   &-0.8\%  &-1.9\%  \\
                         \RMS\  &2.4\%  &1.1\%    &3.2\%   &1.1\%   \\
  \hline
  $HG/LG$: mean                &0.1\%  &-1.9\%   &-1.5\%  &-2.2\%  \\
            \RMS\               &2.0\%  &0.8\%    &2.6\%   &1.2\%   \\
  \hline
   \end{tabular}
     \vspace{3mm}
     
    \caption{Summary of the differences between two calibration schemes, for the calibration coefficients of
       every pixel. Shown are the mean and \RMS\ of the difference distribution over pixels in a camera, given
       in percentage.}

    \label{tab:coeffcomparison}
  \end{center}
\end{table}

\subsection{Verification of the flat-field coefficient calculation}

To check the calculation and merging of flat-field coefficients, these coefficients are recalculated
after application of the merged coefficients to the charge calculation. The newly calculated
coefficients should be close to 1.0 and provide an estimate of the  coefficient uncertainty.

For an example run taken in January the recalculated flat-field coefficients  distributions are
Gaussian and their \RMS\ are $(0.006, 0.006, 0.01, 0.008)$, respectively, for CT1, CT2, CT3 and CT4,
implying an accuracy of $<1$\%.\\

A completely independent estimate of flat-field coefficients can be made using  muon ring images.  As
the Cherenkov emission of single muons is very well understood, it is possible to accurately predict
the charge distribution and shape of  the muon rings observed. For each muon ring, the particular
geome\-tri\-cal parameters (impact distance, arrival angle, \ldots) are fit to the  image. The ratio of the
actual and expected charges gives an estimate of the  efficiency of every pixel on the ring.  This
method is described in detail in \cite{nicolas} and  produces $FF$ coefficients that agree with the
standard calculation at the few  percent level. This level of agreement is encouraging given the
completely different  systematic errors associated with this independent method.

\subsection{Comparison of unusable channels}

Comparisons of unusable channels have also been made between the two calibration techniques.  Except for one or two
channels per camera,  the identified unlocked \ARSs\ are identical. Concerning the other types of problems, the
agreement is generally good, in spite of different methods for unusable channel identification. 

\section{Summary}

Elaborate calibration procedures have been developed and implemented for the four cameras of the \hess\
detector. The calibration algorithms and the monitoring of calibration coefficients monitoring are automated 
and their results are used in the standard data analysis.  Two independent calibration schemes are available
and are in good agreement. The flat-field coefficients have been evaluated by a different method using muon
rings \cite{nicolas} and they are in agreement with the standard coefficients. In conclusion, the
calibration of cameras appears to be robust and accurate.

Pixel amplitudes in units of effective (flat-fielded) photoelectrons should be accurate, for the usual
amplitudes, within a pedestal error of less than 0.1~p.e. and a linear error of less than 5\%, including
systematic uncertainties introduced by the required assumptions about the exact shape of the  amplitude
distributions of single photoelectrons, relevant for the gain calibration. The statistical uncertainties are
negligible compared to the systematic ones because of the high number of calibration runs and the high number
of events.\\

\begin{figure}[h!]
\begin{center}
\psfig{figure=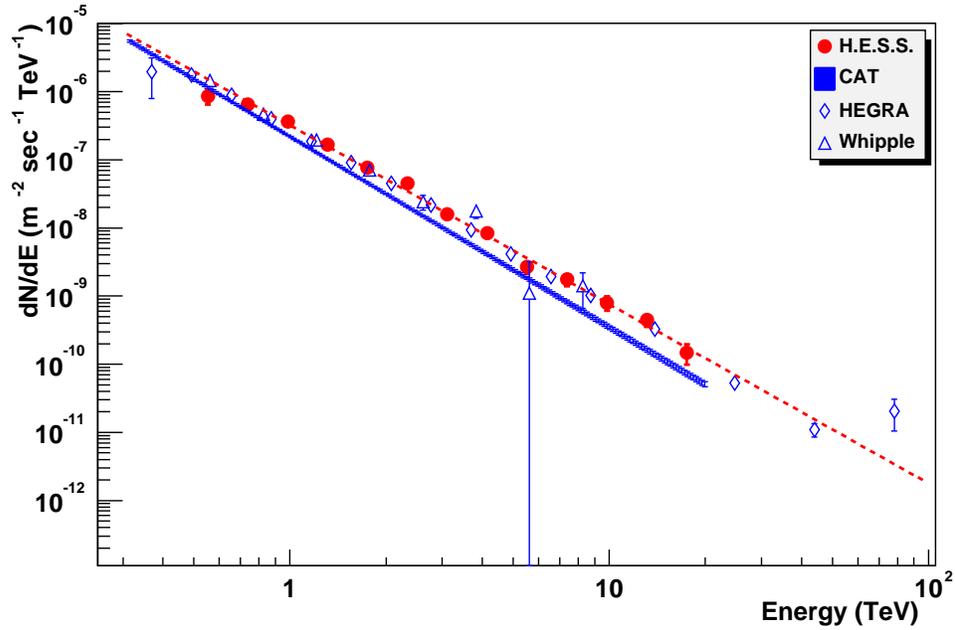,width=1.0\textwidth}
\end{center}
\caption[]{Spectrum of gamma rays from the Crab Nebula, reconstructed on the basis of the
calibration techniques described in this paper, together with results from
CAT \cite{cat_crab}, HEGRA \cite{hegra_crab} and Whipple \cite{whipple_crab}. The dashed line is the
fitted spectrum of the \hess\ data. See \cite{to_be_published} for details.}
\label{fig:crab}
\end{figure}

The camera calibration does not provide the final absolute calibration of the \hess\ detector, which includes the
reflectivity of the mirrors and Winston cones and the absolute quantum efficiency of the \PMTs. The absolute
calibration can either be based on laboratory measurements of these quantities or on the use of muon rings;
the latter approach will be presented in a separate paper. Moreover, the behaviour of the camera trigger has to
be understood correctly in order to estimate the gamma ray detection rates. This is achieved using detailed
simulations of air showers in atmosphere and of the detector, which will be described elsewhere in detail. The
overall quality of the current understanding of the full calibration of the \hess\ instrument is illustrated in
Figure~\ref{fig:crab}, where the flux determined for the Crab Nebula is compared with earlier measurements and
where excellent agreement is found.

\section*{Acknowledgements}

The authors would like to acknowledge the support of their host institutions, and additionally support from
the German Ministry for Education and Research (BMBF), the French Ministry for Research, the Astroparticle
Interdisciplinary Programme of the CNRS, the U.K. Particle Physics and Astronomy Research Council, the IPNP
of the Charles University and the South African Department of Science and Technology and National Research
Foundation. We appreciate the excellent work of the technical support staff in Berlin, Hamburg, Heidelberg,
Palaiseau, Paris, Durham and in Namibia in the construction and operation of the equipment. We acknowledge
contributions by A.~Kohnle and T.~Saitoh during the early stages of the experiment.


\end{document}